\documentclass[runningheads]{llncs}
\usepackage[utf8]{inputenc}

\usepackage[T1]{fontenc}
\usepackage{hyperref}
\usepackage{caption}
\usepackage{subcaption}

\usepackage{xcolor}
\usepackage{amsmath}
\usepackage{cleveref}
\usepackage{graphicx}
\usepackage{color}

\definecolor{orcidlogocol}{HTML}{A6CE39}
\usepackage{tikz}
\usetikzlibrary{svg.path}
\tikzset{
  orcidlogo/.pic={
    \fill[orcidlogocol] svg{M256,128c0,70.7-57.3,128-128,128C57.3,256,0,198.7,0,128C0,57.3,57.3,0,128,0C198.7,0,256,57.3,256,128z};
    \fill[white] svg{M86.3,186.2H70.9V79.1h15.4v48.4V186.2z}
                 svg{M108.9,79.1h41.6c39.6,0,57,28.3,57,53.6c0,27.5-21.5,53.6-56.8,53.6h-41.8V79.1z M124.3,172.4h24.5c34.9,0,42.9-26.5,42.9-39.7c0-21.5-13.7-39.7-43.7-39.7h-23.7V172.4z}
                 svg{M88.7,56.8c0,5.5-4.5,10.1-10.1,10.1c-5.6,0-10.1-4.6-10.1-10.1c0-5.6,4.5-10.1,10.1-10.1C84.2,46.7,88.7,51.3,88.7,56.8z};
  }
}

\renewcommand{\orcidID}[1]{%
  \resizebox{8px}{8px}{
      \href{https://orcid.org/#1}{\tikz[yscale=-1,transform shape]{\pic{orcidlogo}}}}%
}

\begin{document}

\title{Application of Validation Obligations to Security Concerns\thanks{The research presented in this paper has been conducted within the IVOIRE project, which is funded by ``Deutsche Forschungsgemeinschaft'' (DFG) and the Austrian Science Fund (FWF) grant \# I 4744-N. The work of Sebastian Stock and Atif Mashkoor has been partly funded by the LIT Secure and Correct Systems Lab sponsored by the province of Upper Austria.}}
%
%
\author{Sebastian Stock\orcidID{0000-0002-2231-8656} \and
Atif Mashkoor\orcidID{0000-0003-1210-5953} \and
Alexander Egyed\orcidID{0000-0003-3128-5427 }}
\authorrunning{S. Stock et al.}
%
\institute{Johannes Kepler University Linz, Austria\\
\email{firstname.lastname@jku.at}}
\maketitle              
\begin{abstract}
Our lives become increasingly dependent on safety- and security-critical systems, so formal techniques are advocated for engineering such systems. One of such techniques is validation obligations that enable formalizing requirements early in development to ensure their correctness. Furthermore, validation obligations help hold requirements consistent in an evolving model and create assurances about the model's completeness. Although initially proposed for safety properties, this paper shows how the technique of validation obligations enables us to also reason about security concerns through an example from the medical domain.

 \and Formal methods
\end{abstract}

\section{Introduction}
As software systems become more and more responsible for our daily life experiences, it is natural to discuss how to engineer them to ensure their safety and security. Both safety and security are already mature disciplines, and individual processes for dealing with the domain-specific problems are available such as attack trees~\cite{schneier1999attack} for security concerns or model checking~\cite{leuschel2003prob} for safety properties. However, as pointed out in literature (e.g., see~\cite{biro18a,mashkoor20a}), there are a limited amount of cross-cutting techniques available at our disposal, which are capable of being used effectively in both domains.

A validation obligation (VO)~\cite{mashkoor21a} is a logical formula associated with the correctness claim of a given validation property. This technique helps formalize and validate software systems, thus ensuring their overall correctness. While initially proposed for the safety domain, we believe VOs are equally beneficial for validating security concerns in software systems. The benefit of applying VOs to assert the correctness of security concerns lies within the unified approach offered by VOs for modeling all sorts of proprieties, e.g., safety, security, and functional properties. In this fashion, we do not need individual correctness assurance approaches for each set of requirements, which can cause problems while keeping the model consistent. VOs, substantiated with multiple formal techniques, offer a property-agnostic approach for checking completeness and conflict freeness of models.

This paper aims to investigate the application of VOs for the correctness assurance of security concerns. The rest of the paper is structured as follows: \Cref{sec:background} discusses the Event-B method -- the formal method we have used in our approach. Next, \Cref{sec:VOs} introduces the idea of VOs. Then, we exemplify the application of VOs to security concerns in \Cref{sec:example} using an example from the medical domain. Next, \Cref{sec:related} compares the current approaches with VOs for formal modeling of security-critical systems. Finally, we conclude the paper in \Cref{sec:conclusion} with an outlook on future work.

\section{The Event-B method}
\label{sec:background}

State-based formal methods~\cite{mashkoor18a} enable modeling systems with a strict formal syntax, thus allowing to establish correctness assurance with techniques like model checking and theorem proving. This establishes the model's consistency and shows that the model does not lead to a faulty state. Furthermore, state-based formal methods follow the correct-by-construction approach meaning that the model is incrementally enriched with behavior while the correctness of each step is ensured. 

One of the well-known state-based formal methods is Event-B~\cite{abrial2010modeling}. In Event-B states are made up of \texttt{variables} that are constrained by \texttt{invariants}. \texttt{Events} define transitions between the states. A model can be refined via the \texttt{refines} keyword. This means the original specification is advanced concerning variables or events. This refinement relationship, however, needs to be proven to ensure that existing model constraints are not violated in the process of refinement. This is done through proofs. \texttt{contexts} define the static elements of a model and are \texttt{seen} by \texttt{machines} defining the dynamic behavior.

\section{Validation obligations}
\label{sec:VOs}

A VO is composed of a model and a validation task (VT) that must be successfully executed on the model. Once a VT is successfully executed, it establishes the presence of the associated requirement. A VT comes in different forms, e.g., (LTL) model-checking, proof obligations (POs), or even manual inspection of the state space. The input parameters of each VT depend upon the associated requirement. For example, for the LTL model checking, the VT gets an LTL formula as an input parameter. It is the judgment of the designer which VT is best suited to the cause and what are its appropriate parameters. Following is the formal definition of a VO:

\begin{align*}
    VO_{id} = VT_{id}/VT_{context}/VT_{technique}: VT_{parameters}
\end{align*}
As a VO is assigned to a requirement, an \texttt{id} uniquely identfies it. The assigned VT has also an \texttt{id} to identify it. In the area of VOs \texttt{context} refers not to the \texttt{context} of machines as defined earlier but to the context the VO is applied in, i.e., the model we investigate. Second last is the applied \texttt{technique}, e.g., model checking, and last are the input \texttt{parameters}, e.g., the LTL formula.

The concept of VOs should be seen as a carrier technology. It is not bound to a particular formal method. In fact, VOs can be applied to all state-based formal methods alike. In this paper, however, we use Event-B only to exemplify the proposed approach. Following that, we argue that we can validate everything we explicitly model. 

The power of VOs comes from three aspects. 
\begin{enumerate}
    \item By connecting requirements to VOs, designers can rely on the associated VTs for compliance. Once the VO composed of VTs is successfully discharged, it shows the presence of the associated requirement in the model. If the designers change the model subsequently, they can execute the VTs again and assure themselves of the requirement's presence in the evolved model. 
    \item Having all requirements written as VOs lets designers quickly spot conflicting requirements. For example, if requirements are correctly translated into VTs but contradict each other, one of the VOs will fail. In this case, stakeholders can reevaluate the associated requirements. 
    \item VOs open up the ability to view a model from perspectives outside the classical top-down refinement chain. These views can fall into the following categories: (1) abstract views with which we can safely drop information not necessary for reasoning about a VO, and (2) instantiation or scenario views, which transform the abstract model into a concrete example. The output of the associated validation task can give insight into how the view behaves, and the instantiation view can also provide multiple examples of the behavior. Emerging requirements stemming from this exercise can find their way back into the model.
\end{enumerate}

\section{Application of VOs to security concerns}
\label{sec:example}
We now show the application of VOs to establish the correctness of security concerns through the example of a hemodialysis machine model. As aforementioned, we believe that security concerns can be treated like other properties when developing models, which is an advantage as there is no need for their special treatment, thus making the overall modeling process simpler.

\subsection{Illustrative example}
\label{subsec:example}
\begin{figure}[ht!]
	\centering
	\includegraphics[width=0.75\linewidth]{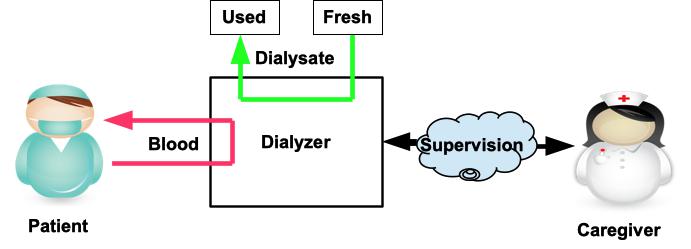}
	\caption{Architecture of hemodialysis machines}
	\label{fig:HDmachine}
\end{figure}

Hemodialysis is a medical treatment that uses a device to clean the blood. The hemodialysis device transports the blood from and to the patient, filters waste and salts from the blood, and regulates the fluid level of the blood. Due to the involved complexity of the dialysis process, the resulted medical treatment is monitored by a professional caregiver for treatment compliance and desired output. Traditionally hemodialysis is performed in a standalone mode, i.e., patients come to a medical facility, get connected to the device, and let dialysis be performed. However, this monitoring is also possible via remote access but demands additional security precautions. The basic architecture of hemodialysis machines is depicted in Figure~\ref{fig:HDmachine}. The requirements specification of hemodialysis machines is discussed in detail in the work of Mashkoor~\cite{mashkoor16a}.

We first design an abstract model and then refine it to show that VOs can be used to gain confidence in the soundness and conflict freeness of the requirements. We then extend the example a second time to show the advantages of creating views on a model. Creating a view means in our context that we keep the abstract model in the background but only focus on one detailed aspect of the model at a time. This helps better understand a model's behavior or debug it by only viewing what is necessary to satisfy a particular requirement.

\subsection{Formal model}
\label{subsec:model}

We start the modeling process with a small subset of the hemodialysis machine requirements specification. We also add some additional security concerns as follows. 
\begin{itemize}
    \item \texttt{SAF1}: In order to start the treatment, the parameters must be within the permissible range.
    \item \texttt{FUN1}: There are three types of staff IDs: maintenance, nurses, and doctors.
    \item \texttt{SEC1}: The staff has to log in to start the treatment.
    \item \texttt{SEC2}: Only doctors and nurses are allowed to start the treatment.
\end{itemize}

\begin{figure}
\centering
\begin{subfigure}{.5\textwidth}
  \centering
  \includegraphics[width=4cm]{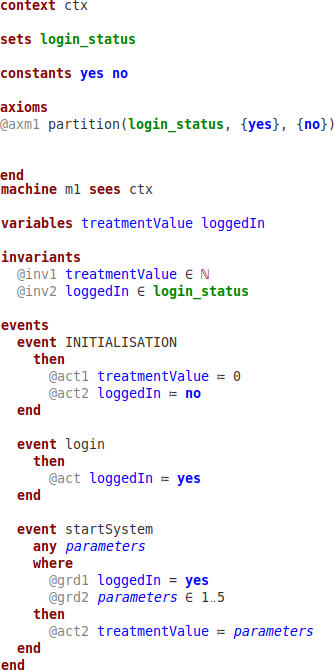}
  \caption{Abstract model}
  \label{fig:baseModel}
\end{subfigure}%
\begin{subfigure}{.5\textwidth}
  \centering
  \includegraphics[width=6cm]{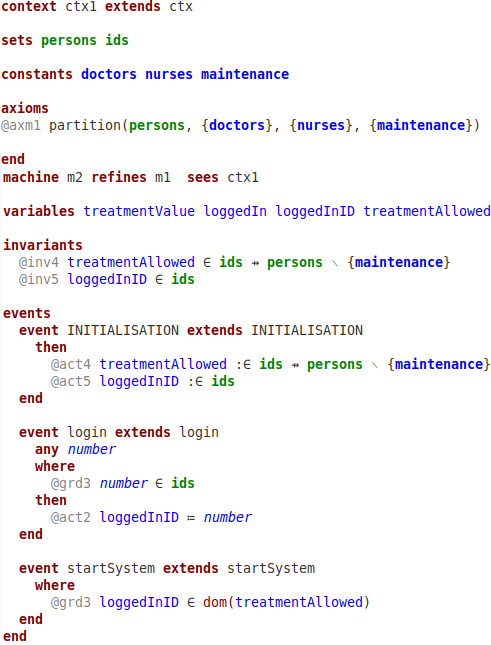}
  \caption{First refinement}
  \label{fig:firstRefinement}
\end{subfigure}
\caption{The formal model}
\end{figure}

\Cref{fig:baseModel} shows the base model of the system. It models the \texttt{SAF1} safety requirement. In the original specification multiple parameters depending on the patients condition can be set thus ensuring a personalized treatment. \texttt{SAF1} is an abstract reference to that, we enter a treatment value which serves as a token for more complex parameters requiered in the original specification. We model \texttt{SAF1} along with the \texttt{SEC1} security requirement. There is an Event-B context \texttt{ctx} modeling the login status. The corresponding machine \texttt{m1} models two events: \texttt{login} and \texttt{startSystem}. Guards, namely \texttt{@grd1} and \texttt{@grd2}, of the \texttt{startSystem} event prevent the event from being fired prematurely. 

\Cref{fig:firstRefinement} shows how the model looks after the refinement. It introduces \texttt{FUN1}  and \texttt{SEC2} requirements into the model. We add roles as data type in the refined context \texttt{ctx1} and configure the set of roles that are allowed to perform a treatment in \texttt{@inv4}. The \texttt{login} event ensures that the role logged into the system is now tracked via \texttt{id}. \texttt{@grd3} in \texttt{startSystem} checks that only allowed \texttt{ids} are able to start the treatment. Also note that the \texttt{startSystem} event in \texttt{m2} extends the event from \texttt{m1} meaning that \texttt{@grd1} and \texttt{@grd2} are still present and active but are hidden form this view of the model to avoid confusion. 

\subsection{Ensuring soundness and conflict freeness of requirements}
\label{sec:soundness}
To each requirement, we associate a VT with which we ensure the presence of this requirement in the model. The VTs can be formulated as:
\begin{itemize}
    \item \texttt{SAF1}/\texttt{m1}/\texttt{PO}: $\texttt{treatmentParameters} \in 1..5$ 
    \item \texttt{SEC1}/\texttt{m1}/\texttt{LTL}:\texttt{G}$(\{\texttt{loggedIn} = \texttt{yes}\} \Rightarrow e(\texttt{startSystem}) )$\footnote{$e(\texttt{startSystem})$ means that this event is enabled because its guard is true.}
\end{itemize}
\texttt{PO} means we have a proof obligation that $\texttt{treatmentParameters} \in 1..5$ is true in every state of the model, meaning the patients treatment parameters are within the allowed range. For the \texttt{LTL} formula \texttt{G}(lobally) means the condition in the brackets is always true. The term inside states that when $\texttt{loggedIn} = \texttt{yes}$ the event \texttt{startSystem} is enabled. \texttt{m1} is the model or in our case the machine we apply the VT to as laid out in \Cref{sec:VOs}.

The proof of the VT for \texttt{SAF1} is relatively simple. We can use the Event-B tool support, i.e., the Rodin platform~\cite{abrial2010rodin}, to show that \texttt{@inv1} holds in every state of the model. The proof can be discharged automatically by the tool. The VT regarding \texttt{SEC1} was established using the LTL formula with the ProB~\cite{leuschel2003prob} model checker, which we employed to do LTL model checking of the model. With both VOs discharged, we can be confident that the concerning requirements are correctly modeled. 

Now we tackle \texttt{FUN1} and \texttt{SEC2}. For those requirements we write the following VTs:
\begin{itemize}
    \item \texttt{FUN1}/\texttt{m2}/PO/\texttt{persons} = $\{\texttt{doctors}, \texttt{nurses}, \texttt{maintenance}\}$
    \item \texttt{SEC2}/\texttt{m2}/LTL/\texttt{G}$(e(\texttt{startSystem}) \Rightarrow \{\texttt{treatmentAllowed}(\texttt{loggedInID}) = \texttt{doctors}\} \vee \{\texttt{treatmentAllowed}(\texttt{loggedInID}) = \texttt{nurses}\} )$
\end{itemize}
We demand proof that the set of roles consists only of doctors, nurses, and maintenance for \texttt{FUN1}. For \texttt{SEC2} we demand that globally the system can only be started if the logged-in role is registered in the set of roles that are allowed to perform treatment, i.e., doctors or nurses.

Again with the help of the corresponding tools, we see that both VTs on the model are executed and satisfied, thus ensuring the soundness of the model \texttt{m2} regarding the requested requirements, i.e., \texttt{FUN1} and \texttt{SEC2}. However, if we run our VTs for \texttt{SEC1} and \texttt{SAF1} previously established on \texttt{m1} on \texttt{m2} to ensure that these requirements are also present in the refinement, we spot a problem. The VT representing \texttt{SEC1} fails. We can find an instance where we are logged into the system but not allowed to start the treatment. This is the case with the login of the maintenance role. 

A failing VT helps us discover a flawed requirement/design in the model, provided that the task was correctly chosen and the parameter was correctly formulated. In our example the requirement encoded as the VT for \texttt{SAF1} is no longer satisfiable as $\texttt{loggedIn} = \texttt{yes}$ is no longer sufficient for starting the system. We could draw three consequences from this: 

\begin{enumerate}
    \item Requirements \texttt{SEC1} and \texttt{SEC2} may be contradicting each other. However, this is is not the case here. 
    \item We could adjust the model by removing the event guard that controls the \texttt{startSystem} event allowing every \texttt{loggedIn} person to start treatment. However, this will lead \texttt{SEC2} to fail; hence the formulation of the requirement as a VO prevented us from introducing new bugs when trying to fix the existing ones. 
    \item Requirements \texttt{SEC1} and \texttt{SEC2} are ambiguous in how they are formulated and need to be clarified. In our case, \texttt{SEC1} is indeed very broad. We can solve the issue by either making \texttt{SEC1} more explicit by stating that the allowed staff has to be logged in or merging the requirement entirely with \texttt{SEC2}.
\end{enumerate}

In our example, we settle for option (3) to clarify requirements: we refactor the corresponding VO, and the fixed VO passes without any new conflict.

\subsection{Creating views}
\label{sec:changingViews}
Different views help designers spot flaws in the model or help non-technical stakeholders better understand the model. Views are a unique form of refinement that aims not to introduce new behavior but to how the model is perceived. VOs help create views by showing their soundness and conflict freeness. However, views affect the associated VT, i.e., the model changes, and the output which decides VT's success or failure also changes accordingly. While the output of a VT is more concrete on an instantiation view, and thus, depending on the task, it can provide a concrete example of why the VT succeeds or fails. The stakeholders can then easily understand the scenario. The knowledge gained from the view can be feedback on the requirements of the model and may result in new VOs to ensure the presence of the emerging requirements.

\begin{figure}
\centering
  \includegraphics[width=6.5cm]{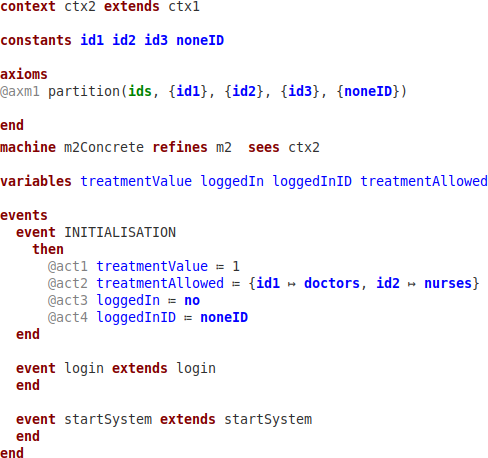}
  \caption{Concrete initialisation for stakeholders}
  \label{fig:concretInit}
\end{figure}

Creating a scenario-view in Event-B is achieved by replacing deferred constructs with concrete ones, thus adapting the model for a specific scenario. Let us consider our hemodialysis example again. \Cref{fig:concretInit} shows a created view. Instead of reusing a deferred function where we map arbitrary \texttt{ids} to roles as we did in \texttt{INITIALISATION}/\texttt{@act4} of \texttt{m2} we use concrete mappings as one can see in \texttt{@act4} of the\texttt{INITIALISATION} of \texttt{m2Concrete}. We successfully execute the VTs and establish conflict freeness and soundness of the requirements for this view. 

\begin{figure}
    \centering
    \includegraphics[width=0.99\linewidth]{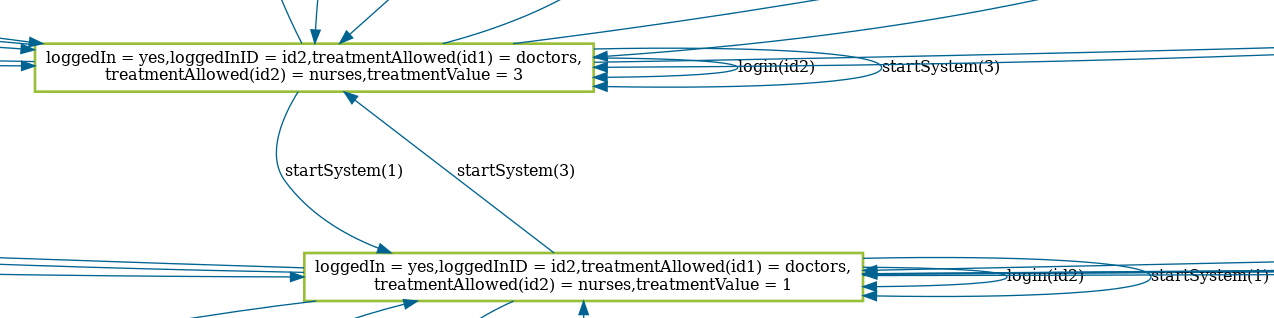}
    \caption{Part of the concrete state-space created with ProB}
    \label{fig:stateSpaceM2}
\end{figure}

The state-space of the model \texttt{m2Concrete} is visualized in~\Cref{fig:stateSpaceM2}\footnote{We omitted parts of the graphic for space reasons.}. It shows two states. One can switch between the states via logging in as a different user or looping back to them by logging in as the same user. From this view, the stakeholder may discover a flaw in the model, for example, the fact that \texttt{startSystem} in the upper state is a loop that always ends in the same state. The consequence of this is that after \texttt{startSystem} was fired by the logged-in role, the logged-in role can be changed. For further treatment, this might be unwanted as the responsible role should not be changed mid-session, nor should the maintenance be able to log in after treatment is started. From this we can formulate a new requirement, for example:
\begin{itemize}
    \item \texttt{FUN2} Once the treatment is started, the logged in ID can not change.
\end{itemize}
This requirement can then be encoded into a VO and run against the original model \texttt{m2}. A VT for this requirement could look like this:
\begin{itemize}
    \item \texttt{FUN2}/\texttt{m2}/\texttt{LTL}:$[$\texttt{startSystem}$]X(G(not(login)))$
\end{itemize}
This formula would check that after the firing of the \texttt{startSystem} event, the \texttt{login} event cannot occur, of course, until the treatment is over.

We can go even a step further. Suppose we find a property of a scenario desirable for all scenarios that are created in the same way, i.e., by using the concrete mapping of \texttt{ids} to roles we used in the example. In that case, we can translate this property into a VO that has to hold for all scenarios that are created, relying on this mapping for initializing the \texttt{treatmentAllowed} variable. Every time we create a scenario view the same way, it must comply with this VO.

\section{Related work}
\label{sec:related}

The Tokeneer case study~\cite{tokeener} is a well-known example of the application of formal methods to the security domain. Tokeener is an extensive system that provides safety for a network of workstations in a protected room, which can only be entered via verified bio-metric scans and id cards. In addition, the Tokeneer system has the task of verifying access for persons and checking certificates for credentials. 
The case study was modeled in different languages like the modeling language Z~\cite{copper2008tokeneer}. For this implementation, there exist several approaches to verify the system, e.g., Cristi{\'a} and Rossi~\cite{cristia2021automatically}. The specification was also translated from Z to Event-B by Rivera et al.~\cite{rivera2016a}. While these contributions show that formal methods can be successfully applied to help design and implement security-critical systems, they only tackle the problem from the verification side. The question of validation is mainly unaddressed.

VOs aim to complement by showing that the actual requirements specified by stakeholders are present in the model. Up to now, many techniques in formal methods have aimed to show the absence of faulty behavior. For example, model checking traverses the state space to check if a state violates a previously defined invariant. Another example is proofs that can show if the access to data structures is well defined. However, verification, i.e., checking whether faulty behavior is absent in the model, is not enough as a designer has no feedback on what the model is capable of. For this, we need validation, and while there exist validation approaches, e.g., proposed by Fitzgerald et al.~\cite{validation_conjectures}, these are tool and language-specific. That is where VOs are a valuable addition, as they provide a tool and language-independent formalism. Requirements are formulated as VOs, which evolve with the model while remaining traceable. Additionally, different model views enable other stakeholders to understand the model and reason about the requirements relevant to them without going into unnecessary details.

\section{Conclusion and future work}
\label{sec:conclusion}
This paper shows how the VO approach can validate security concerns alongside safety and functional properties in a formal model. A uniform approach for validating different system properties is a big plus, thus making the overall development process more straightforward. Furthermore, the VO approach helps spot conflicting requirements and existing bugs and prevents from introducing the new ones. Additionally, we can create views on a model that facilitate the model's understanding by all stakeholders. From views, we can extract knowledge that we can feedback into the model. Finally, we can formulate VOs that describe the scenario's desired outcomes, thus ensuring compliance for the whole development cycle. 

In the future, we want to apply the VO approach to large-scale case studies and show the multitude of VOs in different settings. In such a study, we would like to investigate the following research questions:
\begin{itemize}
     \item What are the limitations of the VO approach?
     \item Is the VO approach equally beneficial to formal methods which are not state-based?
    \item How does the approach scale in large-scale case studies?
\end{itemize}

\bibliographystyle{splncs04}
\bibliography{references}
\end{document}